\documentclass
[prl,superscriptaddress,twocolumn,showpacs,preprintnu
mbers,amsmath,amssymb]{revtex4-1}
\usepackage[dvips]{graphicx}
\usepackage{dcolumn}
\usepackage{amsmath}
\usepackage{ulem}
\usepackage{times}

\def\fun#1#2{\lower3.6pt\vbox{\baselineskip0pt\lineskip.9pt
 \ialign{$\mathsurround=0pt#1\hfil##\hfil$\crcr#2\crcr\sim\crcr}}}
\usepackage{graphicx}
\usepackage{dcolumn}
\usepackage{bm}

\newcommand{\beq}{\begin{equation}}
\newcommand{\eeq}{\end{equation}}
\newcommand{\bea}{\begin{eqnarray}}
\newcommand{\eea}{\end{eqnarray}}

\begin{document}


\title{
Interplay between the $0_2^+$ resonance and the nonresonant
continuum \\ of the drip-line two-neutron halo nucleus $^{22}$C
}


\author{Kazuyuki Ogata}
\email[Electronic address: ]{kazuyuki@rcnp.osaka-u.ac.jp}
\affiliation{%
 Research Center for Nuclear Physics, Osaka University, Ibaraki 567-0047, Japan
}%
\author{Takayuki Myo}%
\affiliation{%
 General Education, Faculty of Engineering, Osaka Institute of Technology,
 Osaka 535-8585, Japan
}%

\author{Takenori Furumoto}
\affiliation{%
 RIKEN Nishina Center, Hirosawa 2-1, Wako 351-0198, Japan
}%

\author{Takuma Matsumoto}
\author{Masanobu Yahiro}
\affiliation{
 Department of Physics, Kyushu University, Fukuoka 812-8581, Japan
}%

\date{\today}

\begin{abstract}
The breakup cross section (BUX) of $^{22}$C by $^{12}$C at 250~MeV/nucleon
is evaluated by the continuum-discretized coupled-channels method
incorporating the cluster-orbital shell model (COSM) wave functions.
Contributions of the low-lying $0_2^+$ and $2_1^+$ resonances
predicted by COSM to the BUX are investigated. The $2_1^+$ resonance
gives a narrow peak in the BUX, as in usual resonant reactions,
whereas the $0_2^+$ resonant cross section has a peculiar shape due
to the coupling to the nonresonant continuum, i.e., the Fano effect.
By changing the scattering angle of $^{22}$C after the breakup,
a variety of shapes of the $0_2^+$ resonant cross sections
is obtained.
Mechanism of the appearance of the sizable Fano effect in
the breakup of $^{22}$C is discussed.
\end{abstract}

\maketitle


{\it Introduction.}
Exploring the frontier of the nuclear chart is one of the most
important subjects in nuclear physics. Properties of neutron
drip-line nuclei, e.g., $^{11}$Li, $^{19}$B, and $^{22}$C, are
therefore crucial for this purpose. Very recently, evidence for
an unbound ground state of $^{26}$O was reported~\cite{Lun12},
which could extend the concept of drip-line nuclei to the
unbound-state regions. In this situation, clarification of
unbound states, i.e., resonance structures, of
nuclei around the neutron drip-line will be a fascinating subject.

Another important aspect of this subject is the {\it figure} of
the cross section to a resonance state. It is well known that
resonant cross sections can have different shapes
from the standard Breit-Wigner form because of the coupling to the
nonresonant continuum.
This is called the background-phase effect or
the Fano effect~\cite{Fan61}.
There have been many examples of the Fano
effect in various research fields, e.g.,
neutron scattering~\cite{Ada49},
Raman scattering~\cite{Cer73},
hypernucleus formation~\cite{MY88},
optical absorption~\cite{Fai97},
and
quantum transport in a mesoscopic system~\cite{Kob02}.
It will be very interesting to
investigate this peculiar effect through
the breakup of neutron drip-line nuclei,
in which the nonresonant continuum is known to play
crucial roles.

In this study we focus on $^{22}$C, the drip-line nucleus of
carbon isotopes.
By measuring the reaction cross section~\cite{Tan10} and the neutron removal
cross section~\cite{Kob11},
ground state properties of $^{22}$C have been intensively studied
so far;
the results strongly support the picture that $^{22}$C is an
s-wave two-neutron halo nucleus, in consistent with the theoretical
prediction of Ref.~\cite{Hor06} based on a $^{20}$C$+n+n$
three-body model.
The dominance of the s-wave configuration of the valence two neutrons
gives large transition probability to the low-energy
nonresonant $0^+$ continuum of $^{22}$C as discussed below.
We can thus expect a significant interference between a possible
low-lying resonance and the nonresonant continuum in the $0^+$ state,
which makes a remarkable change in the shape of the resonant breakup
cross section of $^{22}$C.

In this Letter, we investigate the resonance structure of
$^{22}$C with the three-body
cluster-orbital shell model (COSM)~\cite{SI88}
through the
breakup cross section of $^{22}$C by $^{12}$C at 250~MeV/nucleon
evaluated by the continuum-discretized coupled-channels method
(CDCC)~\cite{Kam86,Aus87,Yah12}.
COSM is a powerful method to describe a system consisting of
a core plus valence nucleons; it has successfully been applied to
studies of the ground and resonance states of $^6$He, $^7$He,
and $^8$He~\cite{Myo07,Myo09,Myo10}. One of the most important
advantages of COSM is
that the relative wave function between the
core nucleus and each nucleon is directly obtained covering a quite
wide space for radial behavior.
CDCC is a sophisticated reaction
model that has been applied to various breakup processes with high success.
Our main purpose is to
investigate how the resonance states of $^{22}$C predicted by COSM
are {\lq\lq}observed'' in the breakup cross section, from the
viewpoint of the Fano effect mentioned above.

{\it Formalism.}
In the present COSM calculation, a $^{20}$C$+n+n$ three-body model
is adopted for the $^{22}$C wave function:
\beq
\Phi_{IM_{I}}^c(  \bm{\eta}_{1},\bm{\eta}_{2})
=
\!\!\!\!
\sum_{\substack{l_{1}j_{1}l_{2}j_{2}\\ i_{1}i_{2}}}
\!\!\!
d_{l_{1}j_{1}l_{2}j_{2}}^{cI,i_{1}i_{2}}
\mathcal{\hat{A}}
\left[
\phi_{l_{1}j_{1}}^{b_{i_{1}}}(\bm{\eta}_{1})
\otimes
\phi_{l_{2}j_{2}}^{b_{i_{2}}}(\bm{\eta}_{2})
\right]  _{IM_{I}},
\label{wf}
\eeq
where $I$ and $M_I$ are the total spin of $^{22}$C and its
third component, respectively, and
${\bm \eta}_i$ ($i=1$ or 2) is the relative coordinate of the $i$th neutron
to the center of the $^{20}$C core.
$\mathcal{\hat{A}}$ represents the antisymmetrization operator for
the two valence neutrons; antisymmetrization between a valence neutron and
a neutron in $^{20}$C is approximately taken into account with the
orthogonal condition model~\cite{Sai69}.
In Eq.~(\ref{wf}), $\phi$ is the Gaussian basis function
\beq
\phi_{ljm_{j}}^{b_{i}}\left(\bm{\eta}\right)
=
\varphi_{l}^{b_{i}}\left(  \eta\right)
\left[
Y_{l}\left(\bm{\hat{\eta}}\right)  \otimes \xi_{1/2}
\right]  _{jm_{j}},
\eeq
where $\xi$ is the spin $1/2$ wave function of neutron and
\beq
\varphi_{l}^{b_{i}}\left(  \eta\right)
=
\sqrt{\frac
{2}{\Gamma\left(  l+3/2\right)  }}
\frac{1}{b_{i}^{l+3/2}}
\eta^{l}\exp\left(  -\frac{\eta^{2}}{2b_{i}^{2}%
}\right)
\eeq
with $\Gamma$ the Gamma function. The range parameters
$b_i$ ($i=1$--$i_{\rm max}$)
are chosen to lie in a geometric progression:
$
b_i = b_1 \gamma^{i-1}.
$
By diagonalizing an internal Hamiltonian $h$
of $^{22}$C with the basis functions,
one obtains eigenstates,
each of which is characterized by
$I$, $M_I$, and the energy index $c$, with the expansion coefficients
$d_{l_{1}j_{1}l_{2}j_{2}}^{cI,i_{1}i_{2}}$.
In the present case, there is only one bound state in $I=0$.
All the other states are located
above the $^{20}$C$+n+n$ three-body threshold,
which are called pseudostates (PS).

Since COSM describes the $^{22}$C wave function covering a quite large
model space,
the PS can be regarded to a good approximation as discretized
continuum states. Then the total wave function of the
$^{20}$C$+n+n+{}^{12}$C four-body reaction system
with the total angular momentum $J$ and its third component $M$
can be expanded as
\beq
\Psi_{JM}(\bm{\eta}_1,\bm{\eta}_2,\bm{R})
=
\sum_{cIL}
\left[
\Phi_{I}^c \left(  \bm{\eta}_{1},\bm{\eta}_{2}\right)
\otimes
\chi_{cIL} \left(  \bm{R} \right)
\right]_{JM},
\label{cdccwf}
\eeq
where
$\chi_{cIL} \left(  \bm{R} \right)$ is the scattering wave of $^{22}$C
in the $(c,I)$ state relative to $^{12}$C;
$L$ ($\bm{R}$) is the corresponding relative angular momentum (coordinate).

By solving the four-body Schr\"odinger equation
\bea
\left[
H-E
\right]
\Psi_{JM}&&(\bm{\eta}_1,\bm{\eta}_2,\bm{R})
=0,
\\
H=
T_{\bm R}+U_{n_1}({\bm R}_1)&&+U_{n_2}({\bm R}_2)+U_{\rm c}({\bm R}_{\rm c})
+h
\label{h4b}
\eea
with the standard boundary condition of
$\chi_{cIL} \left(  \bm{R} \right)$, one may obtain the scattering matrix
to the $(c,I,L)$ channel.
In Eq.~(\ref{h4b}), $T_{\bm R}$ is the kinetic energy operator associated
with $\bm R$, $U_{n_i}$ ($i=1$ or 2) is the neutron distorting potential,
and $U_{\rm c}$ is the potential between the $^{20}$C core and $^{12}$C.
This framework is {\it four-body CDCC}~\cite{Mat04,Man09}
incorporating the COSM wave functions, which we call
{\it COSM-CDCC} below.
We further adopt the prescription~\cite{Mat10}
based on the complex-scaling method (CSM)~\cite{Agu71},
the CSM smoothing method,
to obtain a smooth breakup cross section
$d^2 \sigma/{(d\epsilon d\Omega)}$, i.e.,
the double differential breakup cross section (DDBUX).
Here, $\epsilon$ is the
breakup energy of the $^{20}$C$+n+n$ system measured from the
three-body threshold and $\Omega$ is the solid angle of
the center-of-mass (c.m.) of $^{22}$C after the breakup;
the corresponding polar angle is denoted by $\theta$ below.
It should be noted that in Refs.~\cite{Myo09,Myo10,Mat10}
a continuum level density was shown to be correctly described
in terms of the complex-scaled Green function, which
enables one to obtain continuous strength
functions of physics quantities using the CSM.

{\it Numerical input.}
In the $^{20}$C$+n+n$ three-body Hamiltonian $h$,
we adopt the Minnesota nucleon-nucleon
interaction~\cite{MN} and a Woods-Saxon potential for
the $n$-$^{20}$C system, consisting of the central and spin-orbit parts.
As for the latter, we use Set B parameters of Ref.~\cite{Hor06};
we have slightly changed $V_1$ and $V_s$ to 20.00~MeV and 10.50~MeV,
respectively, so that the 1s state is unbound.
In the COSM calculation, we include the single-particle
configuration of each $n$ up to $l=5$ ($l=4$) for the $0^+$ ($2^+$)
state of $^{22}$C.
The radial wave function between $n$ and $^{20}$C
in each single-particle orbit
is described by 10 Gaussian basis functions;
we use the range parameters of
$b_1=0.3$~fm and $\gamma=1.5$.
We assume a sub-closed shell up to 1d5/2 for neutron in
the $^{20}$C core.

As a result of diagonalization of $h$,
we obtain the $0^+$ ground state at 289~keV below the $^{20}$C$+n+n$
threshold, which is consistent with the empirical
value $420\pm 940$~keV~\cite{Aud03}, together with 604 (1,385) PS
above the threshold in the $0^+$ ($2^+$) state.
The matter radius of the ground state is found to be 3.49~fm.
In the CDCC calculation,
we include the ground state and the
77 (164) PS for $0^+$ ($2^+$)
below $\epsilon=10$~MeV,
which gives a convergence of
the results shown below.

As for the distorting potentials of $n$-$^{12}$C and $^{20}$C-$^{12}$C,
we adopt microscopic single and double folding models, respectively,
with the CEG07b nucleon-nucleon $G$-matrix interaction including the
medium effects \cite{Fur08}.
We use the nuclear densities of $^{12}$C and $^{20}$C given in
Refs.~\cite{Neg70} and \cite{Cha02}, respectively,
with a slight change in the parameters for the former.
CDCC equations between $^{22}$C and $^{12}$C are solved up to $R=30$~fm
and the number of the partial waves is set
to 600. In the CDCC calculation, we use the so-called no-recoil
approximation to the $^{20}$C core, as in the previous
study of Ref.~\cite{TB06}; this approximation is considered to be
valid when the mass of the core nucleus is much larger than the valence
particle(s), which is satisfied well in the present case.

In the CSM smoothing method, we adopt the complex-scaling angle
$\theta_{\rm CSM}$
of $14^\circ$. The basis functions used in diagonalization of
the scaled Hamiltonian $h^{\theta_{\rm CSM}}$ are similar to
above, except that we need finer and wider bases.
We use $(i_{\rm max},b_1,\gamma)=(25,0.2,1.3)$, $(20,0.2,1.3)$, and
$(15,0.3,1.4)$ for the s, d, and other orbits of neutron, respectively.
%
\begin{figure}[b]
\begin{center}
\includegraphics[width=0.45\textwidth,clip]{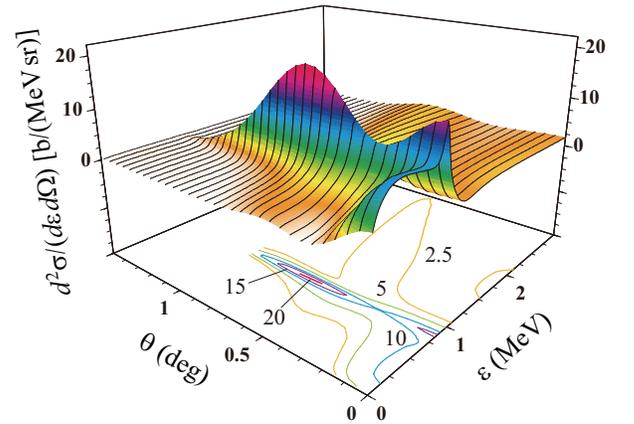}
\caption{(Color online) Double differential breakup cross section (DDBUX)
of $^{22}$C by $^{12}$C at 250~MeV/nucleon.}
\label{ddbux}
\end{center}
\end{figure}

{\it Results and discussion.}
Figure~\ref{ddbux} shows the DDBUX $d^2 \sigma/{(d\epsilon d\Omega)}$
 of $^{22}$C by $^{12}$C at
250~MeV/nucleon calculated by COSM-CDCC. One sees some structures
in the DDBUX, expected to reflect properties of the resonance and
the nonresonant continuum of $^{22}$C.
In fact, COSM predicts some resonance states of
$^{22}$C and $^{21}$C in the energy region shown in Fig.~\ref{ddbux};
the results are summarized in Table~\ref{energy}. The next question
is thus how these resonances contribute to the DDBUX.
%
\begin{table}[hptb]
\caption{Resonance energy $E_{\rm r}$ and width $\Gamma_{\rm r}$ of
$^{22}$C and $^{21}$C.}
\label{energy}
\begin{tabular}{l|cccl}
\hline
\hline
nucleus  & $I^\pi$ & $E_{\rm r}$~(MeV) & $\Gamma_{\rm r}$~(MeV)
& main configuration \\ \hline
$^{22}$C & $0_2^+$ & 1.02 & 0.52 & $(0{\rm d}3/2)^2$ \\
         & $2_1^+$ & 0.86 & 0.10 & $(1{\rm s}1/2)(0{\rm d}3/2)$ \\
         & $2_2^+$ & 1.80 & 0.26 & $(0{\rm d}3/2)^2$ \\
$^{21}$C & $3/2^+$ & 1.10 & 0.10 & $(0{\rm d}3/2)$ \\
\hline
\hline
\end{tabular}
\end{table}

\begin{figure}[htpb]
\begin{center}
\includegraphics[width=0.37\textwidth,clip]{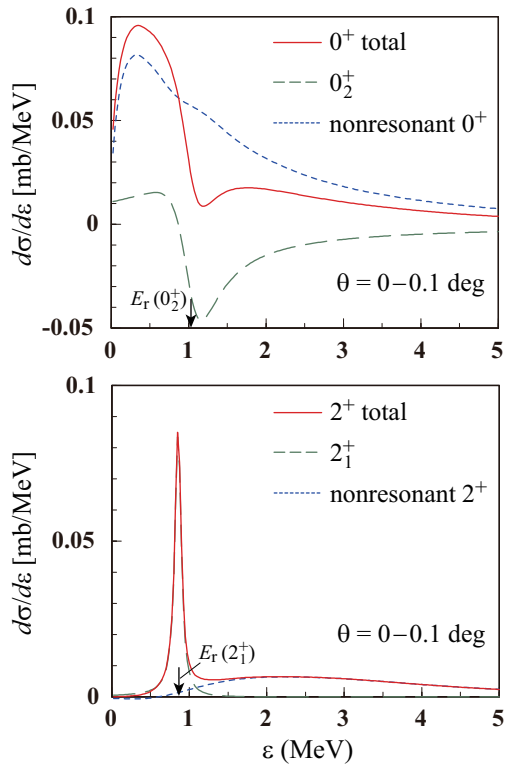}
\caption{(Color online)
Breakup energy distribution corresponding to
$\theta=0^\circ$--$0.1^\circ$ and its
resonant and nonresonant components. See the text for details.
}
\label{decomp1}
\end{center}
\end{figure}

As a great advantage of the CSM-smoothing method,
one can decompose the DDBUX into the components due to the three-body
resonances (each of the $0_2^+$, $2_1^+$, and $2_2^+$ states),
the binary resonance of $^{21}$C coupled with another neutron,
and the nonresonant three-body
continuum. Figure~\ref{decomp1} shows the result of the
decomposition of the breakup energy distribution $d \sigma/{d\epsilon}$,
which is obtained by integrating the DDBUX over $\theta$
from $0^\circ$ to $0.1^\circ$.
%
The upper and lower panels correspond to
the $0^+$ and $2^+$ states of $^{22}$C,
respectively. In each panel, the solid (dotted) line shows the total
breakup cross section (contribution of the three-body nonresonant continuum).
The contribution of the three-body resonance, $0_2^+$ ($2_1^+$) in the
upper (lower) panel, is denoted by the dashed line. In both $I^\pi$ states,
it is found that the contributions from the $^{21}$C binary resonance
are negligibly small. Similarly, the $2_2^+$ resonance gives
an inappreciable cross section.

For the $2^+$ state, one clearly sees that the peak in $d \sigma/{d\epsilon}$
is due to the $2_1^+$ resonance, which has the standard Breit-Wigner form.
It is found that the peak around
$(\epsilon,\theta)=(0.85,0.65)$ shown in Fig.~\ref{ddbux}
 is also due to the $2_1^+$ resonance.
This finding is consistent with the angular dependence of
a BUX corresponding to the multipolarity $\lambda$ of 2, i.e.,
the BUX has a maximum at nonzero $\theta$. Here, $\lambda$ denotes
the multipolarity of $U_{n_i}$ ($i=1$ or 2).
The breakup to the $2^+$ state can thus be regarded as
a standard transition process to a resonance quite isolated from
the nonresonant continuum.

On the other hand, as shown in the
upper panel of Fig.~\ref{decomp1},
the $0_2^+$ resonance has a peculiar form due to the Fano effect.
%
\begin{figure}[htpb]
\begin{center}
\includegraphics[width=0.37\textwidth,clip]{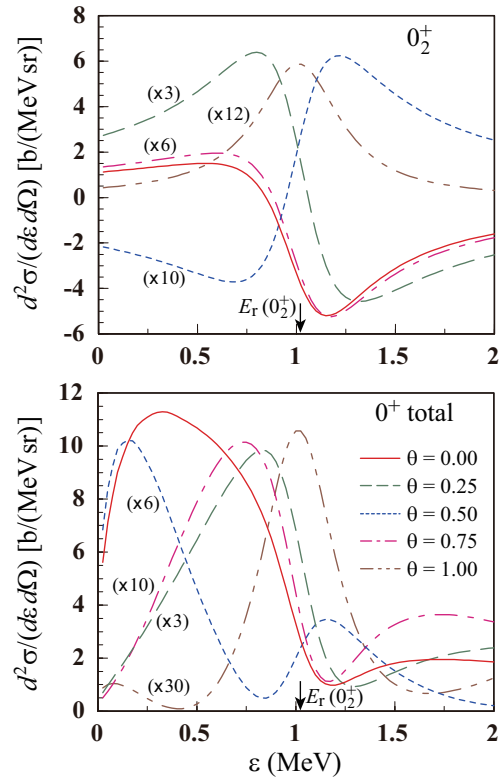}
\caption{
(Color online)
The $0_2^+$ DDBUX (upper panel) and the total
$0^+$ DDBUX (lower panel).
All results except for $\theta=0.00^\circ$ are multiplied by
the numbers beside the lines.
}
\label{decomp2}
\end{center}
\end{figure}
Figure~\ref{decomp2} shows the $\theta$ dependence of
the Fano effect.
The upper and lower panels correspond to the
$0_2^+$ contribution to the DDBUX and the total $0^+$ DDBUX,
respectively.
In each panel, five lines corresponding to
$\theta=0.0^\circ$--$1.0^\circ$ with the increment of $0.25^\circ$ are shown.
One sees distinguished shapes of the $0_2^+$ contribution and the
corresponding $0^+$ total cross section;
we have a variety of shapes of the $0_2^+$ resonant
cross section by changing $\theta$.
Thus, the breakup reaction of $^{22}$C is found to be a {\it site}
in which a sizable Fano effect with various patterns appears.

One of the most important characteristics of $^{22}$C
is the dominance of the $(1{\rm s}1/2)^2$
configuration ($\sim 80$\%) in its ground state.
This gives a large breakup cross section to the low-energy
$0^+$ nonresonant continuum, for which
only the monopole ($\lambda=0$) transition is responsible,
with the same two-neutron configuration.
It should be noted that if neutron has a finite value of $l$, it
hardly exists in the low-energy nonresonant continuum of $^{22}$C
because of the centrifugal barrier.
At the same time, the small but non-negligible $(0{\rm d}3/2)^2$
configuration of about 13\% in the ground state of $^{22}$C
brings the low-lying  $0_2^+$ resonance. This
is essentially due to the closely-located (1s1/2) and (0d3/2)
neutron single-particle orbits of $^{22}$C.
Thus, the resonant
and nonresonant states with the same spin-parity ($0^+$)
strongly affect each other. This is the main reason for the sizable
Fano effect found.
The coexistence of the $0^+$ resonance and nonresonant continuum
will rarely be realized when a core plus one neutron system is
considered; an s-wave neutron cannot form a resonance, except through a
compound process or a Feshbach resonance~\cite{Fes58}. Therefore,
the features of the resonant cross section
shown in the present study are expected to be quite unique to
an s-wave two-neutron halo nucleus, i.e., $^{22}$C.

It should be noted that in the present study we have no so-called
momentum matching condition,
which usually dictates whether or not a resonance is observed
when the reaction $Q$ value is large.
Moreover, since we have only one threshold,
peculiar behavior of a resonance state often found in hadron
physics, e.g., $f_0(980)$~\cite{Pen06}, is not expected.
The present result is purely due to the interplay between
the $0_2^+$ resonance and the $0^+$ nonresonant continuum.

Experimental data of the DDBUX of $^{22}$C are highly desirable
to validate the interesting behavior of the $0^+$ breakup
cross section suggested here.
For this purpose, one must eliminate the $2^+$ cross section from
the total DDBUX.
This can be performed quite easily, because the $2^+$ contribution
will be described well by a standard Breit-Wigner form.
To do this, however, we need experimental data with high energy
resolution;
they can hopefully be obtained at RIBF with utilizing the brand-new
SAMURAI spectrometer.

{\it Summary.}
We have proposed a new framework of four-body CDCC adopting
COSM wave functions, {\it COSM-CDCC}, and applied it to the
breakup process of $^{22}$C by $^{12}$C at 250~MeV/nucleon.
We showed the $2_1^+$ resonance gives a clear peak in the DDBUX,
whereas the $0_2^+$ resonant cross section has a remarkably different
shape from the Breit-Wigner form. The latter is due to the coupling
between the $0_2^+$ resonance and the
$0^+$ nonresonant continuum, i.e., the Fano effect.
The shape of the $0_2^+$ cross section changes drastically with
the scattering angle of the c.m. of the $^{20}$C$+n+n$ system.
The distinguished Fano effect found in the present study
is expected to be unique to an s-wave two-neutron halo nucleus,
i.e., $^{22}$C.

Experimental clarification of the sizable Fano effect on the $0_2^+$
resonance will be very interesting. From the theoretical
side, inclusion of the recoil of the core nucleus $^{20}$C
and its dynamical excitation during the breakup of $^{22}$C
will be important future work.
Extension of COSM-CDCC to five- and six-body breakup reaction
will be a very challenging subject of nuclear reaction studies.

K.~O. thanks Y.~Kikuchi, K.~Mizuyama, T.~Fukui, and H.~Kamano
for valuable discussions.
T.~F. is supported by the Special Postdoctoral Researcher Program of RIKEN.
This research was supported in part by Grant-in-Aid of the Japan
Society for the Promotion of Science (JSPS).

\nocite{*}




\begin{thebibliography}{99}

\bibitem{Lun12}
E. Lunderberg {\it et al.},
Phys. Rev. Lett. \textbf{108}, 142503 (2012).

\bibitem{Fan61}
U.~Fano,
Phys. Rev. \textbf{124}, 1866 (1961).

\bibitem{Ada49}
R.~K.~Adair, C.~K.~Bockelman, and R.~E.~Peterson,
Phys. Rev. \textbf{76}, 308 (1949).

\bibitem{Cer73}
F.~Cerdeira, T.~A.~Fjeldly, and M.~Cardona,
Phys. Rev. B \textbf{8}, 4734 (1973).

\bibitem{MY88}
O.~Morimatsu and K.~Yazaki,
Nucl. Phys. {\bf A483}, 493 (1988).

\bibitem{Fai97}
J.~Faist {\it et al.},
Nature {\bf 390}, 589 (1997).

\bibitem{Kob02}
K.~Kobayashi, H.~Aikawa, S.~Katsumoto, and Y.~Iye,
Phys. Rev. Lett. \textbf{88}, 256806 (2002).

\bibitem{Tan10}
K.~Tanaka {\it et al.},
Phys. Rev. Lett. \textbf{104}, 062701 (2010).

\bibitem{Kob11}
N. Kobayashi {\it et al.}, arXiv:1111.7196 (nucl-ex).

\bibitem{Hor06}
W. Horiuchi and Y. Suzuki,
Phys. Rev. C \textbf{74}, 034311 (2006).

\bibitem{SI88}
Y.~Suzuki and K.~Ikeda,
Phys. Rev. C \textbf{38}, 410 (1988).

\bibitem{Kam86}
M. Kamimura {\it et al.},
Prog. Theor. Phys. Suppl. No.~89, 1 (1986).

\bibitem{Aus87}
N.~Austern {\it et al.},
Phys. Rep. \textbf{154}, 125 (1987).

\bibitem{Yah12}
M.~Yahiro, K.~Ogata, T.~Matsumoto, and K.~Minomo,
Prog. Theor. Exp. Phys. {\bf 1}, 01A209 (2012);
arXiv:1203.5392 (2012).


\bibitem{Myo07}
T.~Myo, K.~Kat\=o, and K.~Ikeda,
Phys. Rev. C \textbf{76}, 054309 (2007).

\bibitem{Myo09}
T.~Myo, R.~Ando, and K.~Kat\=o,
Phys. Rev. C \textbf{80}, 014315 (2009).

\bibitem{Myo10}
T.~Myo, R.~Ando, and K.~Kat\=o,
Phys. Lett. \textbf{B691}, 150 (2010).

\bibitem{Sai69}
S.~Saito,
\newblock Prog. Theor. Phys. {\bf 41} (1969), 705.

\bibitem{Mat04}
T.~Matsumoto {\it et al.},
Phys. Rev. C {\bf 70}, 061601(R) (2004).

\bibitem{Man09}
M.~Rodr\'{i}guez-Gallardo {\it et al.},
\newblock
Phys. Rev. C {\bf 80}, 051601(R) (2009).

\bibitem{Mat10}
T.~Matsumoto, K.~Kat\=o, and M.~Yahiro,
Phys.\ Rev.\ C {\bf 82}, 051602(R) (2010).

\bibitem{Agu71}
J. Aguilar and J. M. Combes,
Commun.~Math.~Phys. {\bf 22}, 269 (1971);
E. Balslev and J. M. Combes,
Commun.~Math.~Phys. {\bf 22}, 280 (1971).

\bibitem{MN}
D.~R.~Thompson, M.~Lemere, and Y.~C.~Tang,
Nucl. Phys. {\bf A286}, 53 (1977).

\bibitem{Aud03}
G.~Audi, A.~H.~Wapstra, and C.~Thibault,
Nucl. Phys. {\bf A729}, 337 (2003).

\bibitem{Fur08}
T.~Furumoto, Y.~Sakuragi, and Y.~Yamamoto,
Phys. Rev. C {\bf 78}, 044610 (2008),
{\it ibid.} {\bf 80}, 044614 (2009).

\bibitem{Neg70}
J.~W.~Negele,
Phys. Rev. C {\bf 1}, 1260 (1970).

\bibitem{Cha02}
L.~C.~Chamon {\it et al.},
Phys. Rev. C {\bf 66}, 014610 (2002).

\bibitem{TB06}
J.~A.~Tostevin and B.~A.~Brown,
Phys. Rev. C {\bf 74}, 064604 (2006).

\bibitem{Fes58}
H.~Feshbach, Ann. Phys. {\bf 5}, 357 (1958),
{\it ibid.} {\bf 19}, 287 (1962).

\bibitem{Pen06}
M.~R.~Pennington, Int. J. Mod. Phys. A, {\bf 21}, 747 (2006).

\end{thebibliography}
\end{document}